# Accelerated Discovery of Vanadium Oxide Compositions: A WGAN-VAE Framework for Materials Design

Danial Ebrahimzadeh, Sarah S. Sharif, Yaser M. Banad

*School of Electrical and Computer Engineering, University of Oklahoma, Norman, OK, USA*

**Abstract**

The discovery of novel materials with tailored electronic properties is crucial for modern device technologies, but time-consuming empirical methods hamper progress. We present an inverse design framework combining an enhanced Wasserstein Generative Adversarial Network (WGAN) with a specialized Variational Autoencoder (VAE) to accelerate the discovery of stable vanadium oxide (V–O) compositions. Our approach features (1) a WGAN with integrated stability constraints and formation energy predictions, enabling direct generation of thermodynamically feasible structures, and (2) a refined VAE capturing atomic positions and lattice parameters while maintaining chemical validity. Applying this framework, we generated 451 unique V–O compositions, with 91 stable and 44 metastable under rigorous thermodynamic criteria. Notably, we uncovered several novel $V_2O_3$ configurations with formation energies below the Materials Project convex hull, revealing previously unknown stable phases. Detailed spin-polarized DFT+U calculations showed distinct electronic behaviors, including promising half-metallic characteristics. Our approach outperforms existing methods in both quality and stability, demonstrating about 20% stability rate under strict criteria compared to earlier benchmarks. Additionally, phonon calculations performed on selected compositions confirm dynamic stability: minor imaginary modes at 0 K likely stem from finite-size effects or known phase transitions, suggesting that these materials remain stable or metastable in practical conditions. These findings establish our framework as a powerful tool for accelerated materials discovery and highlight promising V–O candidates for next-generation electronic devices.

## 1. Introduction

The discovery and development of novel materials are pivotal for advancing modern electronic technologies [1], particularly in applications such as semiconductors, memories, batteries, and photovoltaics [2]. Among various material systems, vanadium oxides (V-O) have emerged as particularly promising candidates due to their diverse electronic properties, phase transitions, and structural versatility. These materials exhibit remarkable characteristics including metal-insulator transitions, varied oxidation states, and tunable electronic properties, making them essential for next-generation electronic devices [3]. However, the systematic discovery of new stable V-O compositions with desired properties remains challenging due to the complex relationship between structure and functionality.

Traditionally, materials exploration has relied heavily on chemical intuition and empirical approaches, such as trial-and-error synthesis and high-throughput screening. While these methods have proven successful in identifying some useful materials, they face significant limitations in efficiency, scalability, and their ability to systematically explore the vast chemical space of possible V-O compositions. The growing demand for materials with precisely tailored properties has created an urgent need for more sophisticated approaches to accelerate materials discovery, particularly those enabling inverse design—where desired properties guide the search for optimal material compositions and structures [4].

High-throughput (HTP) computational workflows, leveraging methods such as Density Functional Theory (DFT), have revolutionized materials discovery by enabling the rapid evaluation of potential candidates. For example, Wines et al utilized an HTP workflow to explore two-dimensional materials exhibiting ferroelectricity and Rashba spin splitting [5]. However, these workflows are computationally intensive and often require significant resources to explore even a small fraction of the possible material space.

To address these challenges, global optimization techniques, including genetic algorithms and Bayesian optimization, have emerged as essential tools. These methods provide efficient navigation through vast and complex design spaces, significantly reducing computational overhead [6]. The CALYPSO method [7] utilizes Particle Swarm Optimization to predict stable and metastable crystal structures from scratch. Its integration of symmetrical constraints and diversity enhancements has proven effective in exploring complex configurational spaces for material discovery. Further developments in this area include Bayesian Optimization approaches for identifying optimal nanoporous materials [8], with Yamashita et al. demonstrating efficient crystal structure prediction through reduced trial numbers in systems like NaCl and $Y_2Co_{17}$ [9].

Evolutionary algorithms have shown particular promise in crystal structure prediction and material design [10]. Notable advances include the COPEX algorithm by Liu et al. [11], which extends the USPEX evolutionary algorithm through co-evolutionary approaches for complex material systems. The work of Oganov and Glass [12] further demonstrated the evolutionary methods when combined with ab initio calculations, enabling reliable identification of stable and metastable structures across diverse materials without requiring experimental input.

The emergence of machine learning, coupled with comprehensive materials databases such as the Materials Project (MP) [13] and Open Quantum Materials Database (OQMD) [14], has transformed the landscape of inverse materials design. Deep learning techniques [15], including neural networks, have shown remarkable success in diverse applications, such as analyzing high-throughput Transmission Electron Microscope data [16]. More notably, generative models like Variational Autoencoders (VAEs) [17] and Generative Adversarial Networks (GANs) [18], alongside Reinforcement Learning and Invertible Neural Networks, have revolutionized inverse design by enabling the generation of novel materials with tailored properties [19]. The neural network (ResNet18) was used to analyze high-throughput Transmission Electron Microscope (TEM) data for phase determination. Raccuglia et al. demonstrated machine learning's potential for predicting lab experiment outcomes [20], and Pan et al. developed a Reinforcement Learning framework for identifying promising inor-ganic compounds while maintaining chemical validity constraints. Their model learns chemical guidelines, such as charge and electronegativity neutrality, to generate novel compounds targeting properties like formation energy [21]. The MatDesINNe framework by Fung et al. introduced an inverse design technique utilizing invertible Neural Networks to establish bidirectional mappings between material design parameters and target properties [22].

While generative models have shown promise in capturing and replicating underlying data distributions, recent advances have expanded their utility to classification tasks [23], and transfer learning approaches [24]. These models have become particularly valuable for inverse design, effectively bridging computational modeling and experimental validation [25][26]. Generative models like VAEs are popular, designing molecules by learning continuous representations of existing materials to generate new compounds with desired properties [27]. Noh et al. developed the iMatGen model, which uses VAEs and sampling techniques to represent image-based crystals and generate distinct $V_xO_y$ compounds [28]. In another work, Boltzmann Generators have been adapted to discover new crystal structures, learning the underlying probability distributions of atomic arrangements in stable crystals [29]. Recent work by Nouira et al. with CrystalGAN generates ternary compositions from binary structures using a vector-based representation [30]. Teng Long et al. have further demonstrated the potential of generative models in materials discovery, proposing the Deep Convolutional GAN model to generate new binary materials with specific properties [31].

Despite these advances, two critical challenges persist in the inverse design of electronic materials. The first challenge relates to chemical constraints, which involve ensuring that generated $V_xO_y$ compounds exhibit realistic thermodynamic stability and oxidation states comparable to known phases. Oxygen vacancies and mixed oxidation states further complicate the stability of these structures. The second challenge pertains to the structural representation of 3D crystal structures, as unlike molecular systems with established representations like SMILES [32], periodic materials require encoding methods that maintain atomic connectivity and lattice parameters to preserve material properties [33].

To address these challenges, we present two key innovations in this study. We developed an enhanced WGAN-based generative framework specifically optimized for V-O systems. Unlike traditional GANs that suffer from mode collapse and gradient vanishing, our model incorporates formation energy constraints to improve thermodynamic stability, leading to more physically meaningful generated structures. This approach enables robust handling of complex structural and chemical constraints, improved training stability through Wasserstein distance optimization, and the integration of formation energy predictions during material generation.

In addition, we developed a refined VAE architecture utilizing voxel-based representation that encodes atomic positions and lattice parameters as a unified 3D grid format. Specifically, separate voxel grids represent the atomic positions of vanadium and oxygen, while an additional grid encodes lattice parameters, ensuring periodicity is maintained. This novel approach effectively encodes both atomic positions and lattice parameters, maintains chemical validity through learned constraints, and creates a continuous, invertible latent space suitable for V-O material generation.

The effectiveness of our integrated approach is demonstrated through extensive validation and discovery. We generated and characterized numerous V-O compositions, including $VO_2$, $V_2O_3$, and $V_2O_5$, with particular success in identifying novel stable structures. Our selection of $VO_2$, $V_2O_3$, and $V_2O_5$ as the primary targets of this generative framework is based on their growing research interest, prevalence in vanadium oxide studies, and significant applications in electronic and energy-related fields. As shown in Fig. 1, the research frequency of these materials in Web of Science has been steadily increasing over the past years, highlighting their continued relevance in materials science. Additionally, their distribution among studies on vanadium oxides demonstrates that they represent a substantial portion of research efforts, particularly in areas related to phase transitions, energy storage, and catalysis. Beyond their academic significance, these materials have diverse and critical applications, including batteries, supercapacitors, memristors, sensors, electrochromic devices, and catalysis, making them technologically important.

Most notably, we discovered several $V_2O_3$ materials with formation energies lower than those in the Materials Project and OQMD convex hulls. These findings not only validate our methodology but also suggest the existence of previously unknown, thermodynamically favorable V-O phases suitable for electronic applications.

In Section 2, we detail our methodological framework, describing both the data preparation and the enhanced generative models used for encoding tasks and material generation.

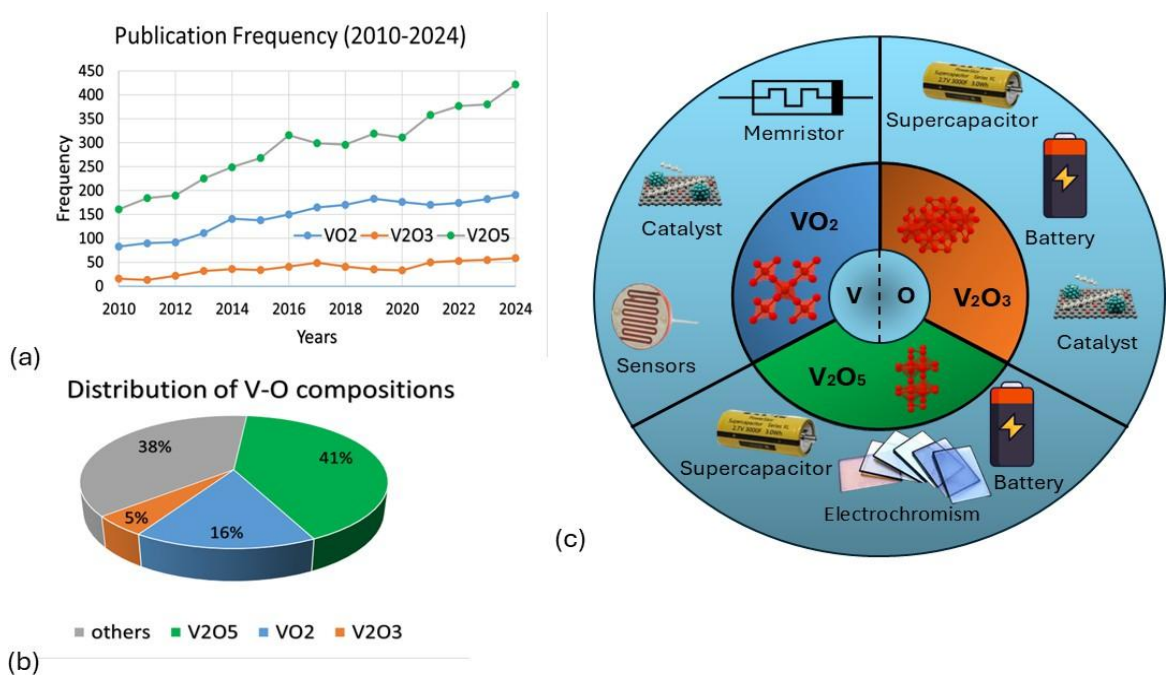

Figure 1: Research trends and applications of vanadium oxide (V-O) compositions. (a) Publication trend analysis showing the increasing research interest in $VO_2$, $V_2O_3$, and $V_2O_5$ from 2010 to 2024, based on Web of Science data. $VO_2$ and $V_2O_5$ demonstrate particularly strong growth trajectories, with $VO_2$ publications increasing from 200 to 750 articles annually. (b) Relative distribution of research focus among V-O compositions, where $V_2O_5$ (41%) and $VO_2$ (16%) account for the majority of targeted studies, while $V_2O_3$ (5%) represents an emerging research direction. The 'Others' category (38%) encompasses various compositions including ternary compounds and alternative stoichiometries. (c) Circular representation of major technological applications, highlighting the versatility of V-O materials across six key domains: electronic memory (memristors), energy storage (supercapacitors and batteries), catalysis, sensing, and electrochromic devices. Each V-O composition shows distinct advantages in specific applications, as indicated by their sectoral placement.

Section 3 presents a comprehensive set of results, including stability analyses, phonon calculations, and electronic structure evaluations for the newly generated V–O compositions. We then discuss our findings on structure-property relationships, as well as the broader implications for inverse materials design. Finally, in Section 4, we summarize the principal conclusions and outline future directions for integrating these techniques into the accelerated discovery of functional materials.

## 2. Method

### 2.1. Inverse Design Framework

Inverse design represents a paradigm shift in materials discovery, where the process begins with desired properties and works backward to identify materials that meet these criteria (Fig. 2). This approach contrasts with traditional forward design methods, where materials are first synthesized or simulated, and their properties are subsequently analyzed. Our implementation of this approach consists of three main components: (1) dataset construction and curation, (2) material representation and encoding, and (3) generative modeling for new material discovery.

### 2.2. Data Generation and Curation

To address the limited availability of known vanadium oxide compositions, which are insufficient for training robust machine learning models, we developed a systematic data generation pipeline. Building on prior work [28] that curated 10,981 binary materials from the

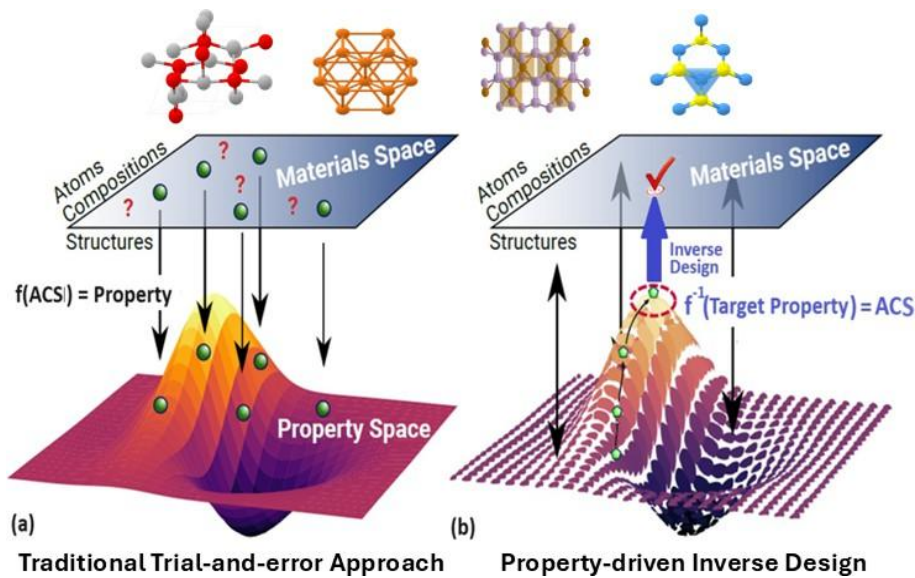

Figure 2: Comparison of traditional trial-and-error materials discovery with property-driven inverse design. (a) In the traditional approach, the material's properties are explored by testing various combinations of atomic compositions, structures, and material spaces, often requiring extensive trial-and-error. (b) The inverse design approach starts with a target property and employs computational models to identify the optimal atomic compositions and structures, streamlining the discovery process and enabling efficient exploration of the materials space.

Materials Project database (excluding structures with excessive unit cells), we implemented a targeted substitution approach specifically for V-O systems and incorporated DFT calculations to compute the formation energies of the V-O structures. This approach yielded a comprehensive dataset capturing diverse compositional and configurational spaces necessary for training robust machine learning models.

## 2.3. Crystal Structure Representation

Traditional crystallographic formats like CIF (Crystallographic Information File) pose challenges for machine learning due to their discontinuous nature and lack of a structured spatial representation suited for deep learning models. We address this limitation through a voxel-based representation approach that has been proven its ability to captures both spatial and compositional information in a format compatible with deep learning architectures that utilize convolutional layers [34].

Voxel grids offer a spatially detailed depiction of three-dimensional objects, making them particularly well-suited for capturing the intricate structural characteristics of crystals. Unlike graph-based approaches, which may struggle with encoding periodicity and lattice interactions, voxels provide a fixed-resolution spatial encoding that preserves atomic connectivity and lattice parameters. The voxelization process allows the model to efficiently learn spatial relationships between atoms, ensuring structural validity in generated materials.

For a typical binary compound, we utilized a method that defines three distinct voxel grids: two grids to represent the spatial distribution of atomic positions for each of the two constituent elements, and a third grid to encode lattice parameters such as dimensions, lengths, and angles. This structured input format helps maintain atomic relationships, enabling stable material generation. Although voxel representation primarily encodes structural information, it indirectly influences electronic properties through structure–property relationships.

## 2.4. Variational Autoencoder Architecture

We developed an enhanced VAE to encode materials into a continuous latent space that seamlessly integrates chemical composition and structural features. The encoder compresses the high-dimensional input data into a compact latent space, while the decoder reconstructs the original input from these latent representations. This continuous and smooth latent space makes the VAE particularly effective for modeling complex material properties.

In our implementation, the VAE was trained using voxelized crystalline structures as input, enabling the model to capture both atomic and lattice-level information. To further enhance the learning process, we incorporated a novel mechanism for handling residual connections, ensuring robust and accurate representations by mitigating potential issues with data misuse. Specifically, in the decoder, we employed residual connections inspired by ResNet [35], where feature maps from earlier layers were upsampled and combined with those in later layers. This architecture significantly improved the model's capacity to learn hierarchical and nuanced material features.

The training process simultaneously optimized the encoder and decoder networks to minimize the total loss, ensuring the latent space effectively preserved both the structural and chemical characteristics of the input data. The total loss function consisted of two key components: the reconstruction loss, which quantifies how accurately the autoencoder reconstructs input data from latent representations, and the Kullback-Leibler (KL) divergence loss, which regularizes the latent space for smoothness and continuity. The reconstruction loss is formulated as:

$$ReconstructionLoss = \frac{1}{N} \sum_{i=1}^{N} \|x_i - \hat{x}_i\|^2 \tag{1}$$

where $x_i$ is the input data, and $\hat{x}_i$ is the reconstructed output from the decoder. The KL divergence encourages the latent variables' distribution to approximate a standard normal distribution. By acting as a regularizer, the KL divergence prevents overfitting to the training data, promoting smoothness and generalizability of the learned latent space. The KL divergence loss is defined as:

$$KLDivergenceLoss = -\frac{1}{2} \sum_{j=1}^{z_{size}} \left(1 + \log(\sigma^2) - \mu_j^2 - \sigma_j^2\right) \tag{2}$$

where $\mu_j$ and $\sigma_j^2$ are the mean of the j-th latent variable produced by the encoder, respectively. The total loss function, which guides the training of the VAE, combines the reconstruction loss and the KL divergence loss as follows:

$$TotalLoss = ReconstructionLoss + \beta \cdot KLDivergenceLoss \tag{3}$$

Here, $\beta$ is a scaling factor that controls the trade-off between the reconstruction accuracy and the regularization imposed by the KL divergence. A higher $\beta$ encourages greater regularization, resulting in a smoother and more structured latent space.

Through this carefully balanced training process, the VAE learns to encode input data into a meaningful and continuous latent space while maintaining the ability to decode it back into its original representation. This reversibility ensures the integrity of the encoded latent representations and provides a powerful framework for material property modeling and design.

In our architecture, the encoder employs 3D convolutional layers with Leaky ReLU activation, batch normalization, and dropout, outputting mean and log variance vectors for reparameterization in the latent space. The decoder mirrors this with 3D transposed convolution layers and residual connections enhanced by custom 3D upsampling, enabling precise and robust reconstruction of input structures. A beta-VAE framework balances reconstruction accuracy and latent space regularization via KL divergence, with KL annealing and learning rate scheduling ensuring stable training. This architecture effectively integrates spatial feature extraction, latent regularization, and reconstruction, making it well-suited for complex lattice design.

### 2.5. Generating New Materials

Generative adversarial networks (GANs) consist of two competing neural networks: the generator ($G$) and the discriminator ($D$), trained simultaneously in an adversarial framework. The generator $G(z; \theta_G)$ maps random noise $z$ into synthetic data samples $G(z)$, while the discriminator $D(x; \theta_D)$ evaluates whether a given sample $x$ originates from the real data distribution ($x \sim P_{\text{data}}$) or is generated ($x = G(z)$). This adversarial setup enables the generator to produce increasingly realistic outputs as training progresses.

The Wasserstein Generative Adversarial Network (WGAN) [36] extends the traditional GAN framework by addressing critical challenges such as unstable training, mode collapse, and vanishing gradients. Instead of a binary discriminator, the Wasserstein GAN (WGAN) employs a "critic" $C$ that assigns a continuous score to measure the realness of samples. WGAN optimizes the Wasserstein distance (also known as Earth Mover's distance) between the real data distribution $P_{\text{data}}$ and the generator's output distribution $P_G$, leading to more stable training dynamics and enhanced generative performance. These advancements make WGAN a powerful tool for generating high-quality data, particularly in applications requiring precision and reliability.

As illustrated in Fig.3, the inverse design process begins with a materials database. These materials are represented in voxel space and subsequently encoded into a latent space using a VAE specially optimized for this task. The VAE extracts essential structural and compositional features from the materials, encoding them into a continuous latent space compatible with WGAN. This structured latent representation ensures that the generator operates within a well-defined space, facilitating the creation of realistic and stable material compositions. The framework now incorporates a comprehensive two-stage stability verification protocol: thermodynamic stability and dynamic stability. The combined use of VAE, WGAN, CNN, and multi-level stability analysis provides a robust framework for inverse design, accelerating the discovery of novel materials with both desirable properties

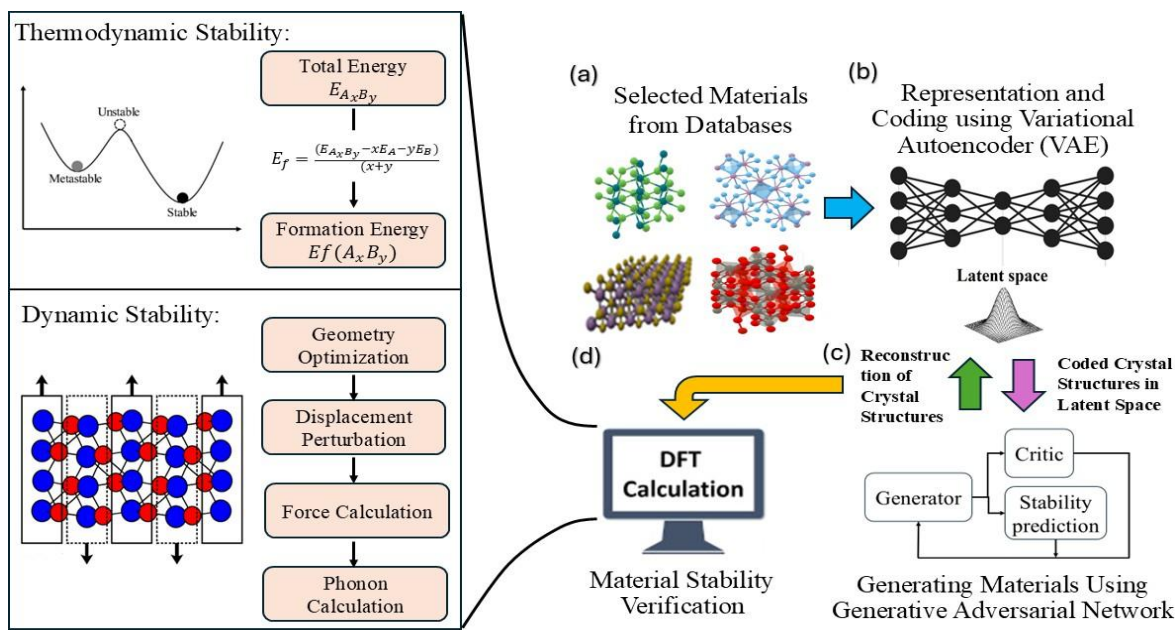

Figure 3: Comprehensive workflow of the inverse design framework for materials discovery: (a) Selection of materials from databases, (b) Representation and encoding using our enhanced VAE into a continuous latent space, (c) Generation of novel materials through the WGAN with stability prediction, and reconstruction of crystal structures, (d) Rigorous stability verification through DFT calculations, incorporating detailed thermodynamic stability assessment (formation energy) and dynamic stability evaluation (phonon calculation).

and practical synthesis potential.

The generator in the proposed architecture employs a sequential design that combines dense layers, upsampling, and convolutional layers, further enhanced with batch normalization. The discriminator, aligned with WGAN principles, employs convolutional layers with strides, Leaky ReLU activations, dropout, and L2 regularization to assess the realism of generated samples, ensuring a stable training process and superior convergence. The composite model combines these elements and leverages a dual loss function: the WGAN loss for ensuring image quality and a custom formation energy loss for enforcing properties. This WGAN-based framework enables the generation of high-quality, domain-relevant data with enhanced stability and realistic features.

## 3. Discussion and Results

### 3.1. Computational Setup and Methodology

The research was conducted using a system equipped with an Intel Core i7 CPU, 8GB of RAM, and an Nvidia 820 GPU. The computational environment was developed using Python 3, employing libraries such as TensorFlow, Keras, and ASE (Atomic Simulation Environment) for materials science operations. To ensure a fair comparison, both GAN and WGAN models were trained under identical conditions for 5,000 iterations.

Table 1 summarizes the key parameters of the model. The 3D voxel representation for atomic sites utilizes a $64 \times 64 \times 64$ grid, while the lattice encoding is represented using a $32 \times 32 \times 32$ grid. The latent vector size for the Lattice VAE is 25, and the latent vector size for the Sites VAE is 200. These two encoded representations are then merged into a new combined representation, which serves as input for the WGAN. So, the latent vector size for WGAN is 200. Moreover, the number of training data is 10981 and we generated 451 new materials that we use DFT for calculating their formation energies.

For stability validation, we performed DFT calculations using the PBE functional [37] and PAW-PBE pseudopotentials [38] for V and O atoms, implemented in the VASP program package [39]. Structural relaxation was performed using the conjugate gradient descent method, with convergence criteria set at $1.0e^{-5}$ for energy and $0.05\,\mathrm{eV/\mathring{A}}$ for force. Computational efficiency was maintained through the use of relatively sparse reciprocal space meshes with a grid spacing of $0.5\,\mathring{A}^{-1}$, combined with a plane-wave cutoff energy of 500 eV. The formation energy was computed as:

$$E_f = \frac{E_{V_xO_y} - (xE_V + yE_O)}{x+y} \tag{4}$$

Table 1: Model parameters and dataset statistics.

| Parameters | voxel site | voxel lattice | vector site VAE | vector lattice VAE | vector WGAN | Training data | Generated data |
|---|---|---|---|---|---|---|---|
| Values | 64×64×64 | 32×32×32 | 200 | 25 | 200 | 10981 | 451 |

where $E_{V_xO_y}$ represents the total energy of the V-O structure, $E_V$ and $E_O$ are the energies of isolated vanadium and oxygen atoms, respectively. Electronic property analysis, including DOS and band structure calculations, was performed using VASP with post-processing through VASPKIT [40].

To assess the dynamic stability of our selected structures, we conducted phonon calculations using a first-principles supercell approach, implemented via the PHONOPY package [41][42] in conjunction with the VASP. Prior to phonon calculations, all structures were fully optimized concerning energy, forces, and stress, ensuring that the residual forces were below $0.01\,\mathrm{eV}\cdot\mathrm{\mathring{A}}^{-1}$, in accordance with the convergence criterion set in our calculations.

*3.2. Model Performance Analysis*

The effectiveness of our VAE implementation is demonstrated in Fig. 4, which shows learning curves for both lattice and site components. The models exhibit rapid initial convergence, with the lattice VAE stabilizing around epoch 50 and the site VAE reaching stability near epoch 15. The close alignment between training and test losses indicates minimal overfitting and strong generalization capabilities. These results validate our VAE architecture's ability to learn effective latent representations for both lattice and site structures while maintaining stable, low loss values throughout training.

The comparative analysis between GAN and WGAN architectures, illustrated in Fig. 5, reveals significant differences in training stability. The traditional GAN's generator loss shows considerable fluctuations throughout the training process and does not stabilize even after 5000 training epochs. This instability reflects common challenges such as vanishing gradients, mode collapse, and generator-discriminator imbalance, which hinder stable learning [43]. In contrast, our WGAN implementation demonstrates remarkably stable convergence, characterized by a smooth, steady decline in generator loss. This enhanced stability can be attributed to the Wasserstein distance metric, which provides a more informative loss function, allowing the generator to receive continuous and meaningful gradient updates. Additionally, weight clipping prevents discriminator dominance, ensuring consistent training stability. These improvements effectively mitigate training instabilities while maintaining robust gradient flow and improving sample diversity. The superior training dynamics of WGAN underscore its suitability for generating diverse and physically meaningful materials in inverse design applications.

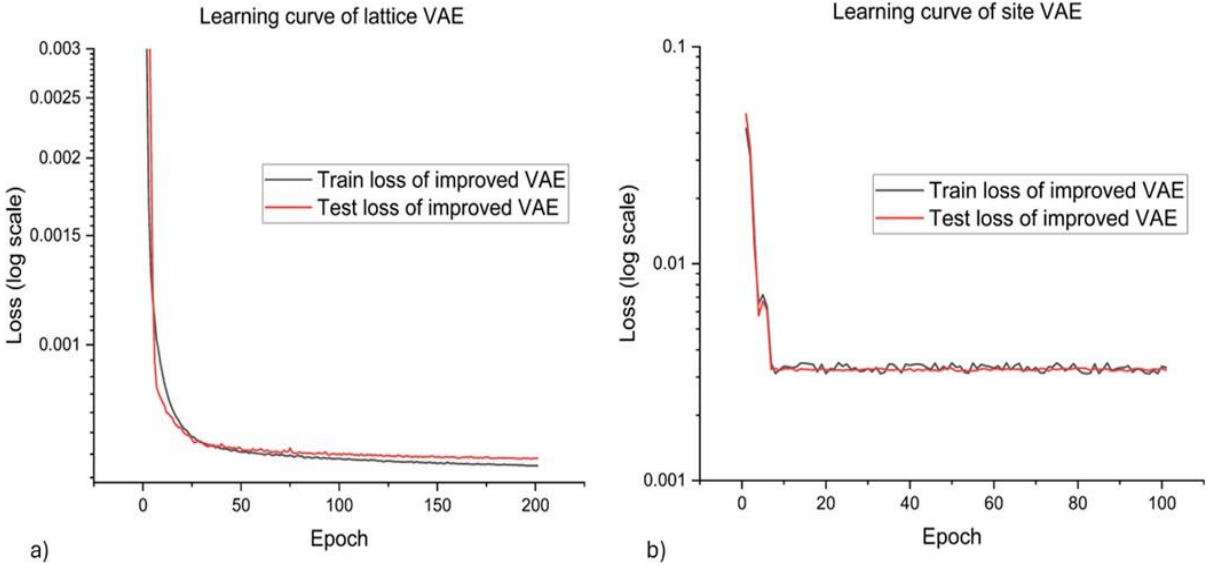

Figure 4: Learning curves of the proposed VAE model. (a) Lattice VAE and (b) Site VAE show the training (black) and test (red) losses over epochs on a logarithmic scale.

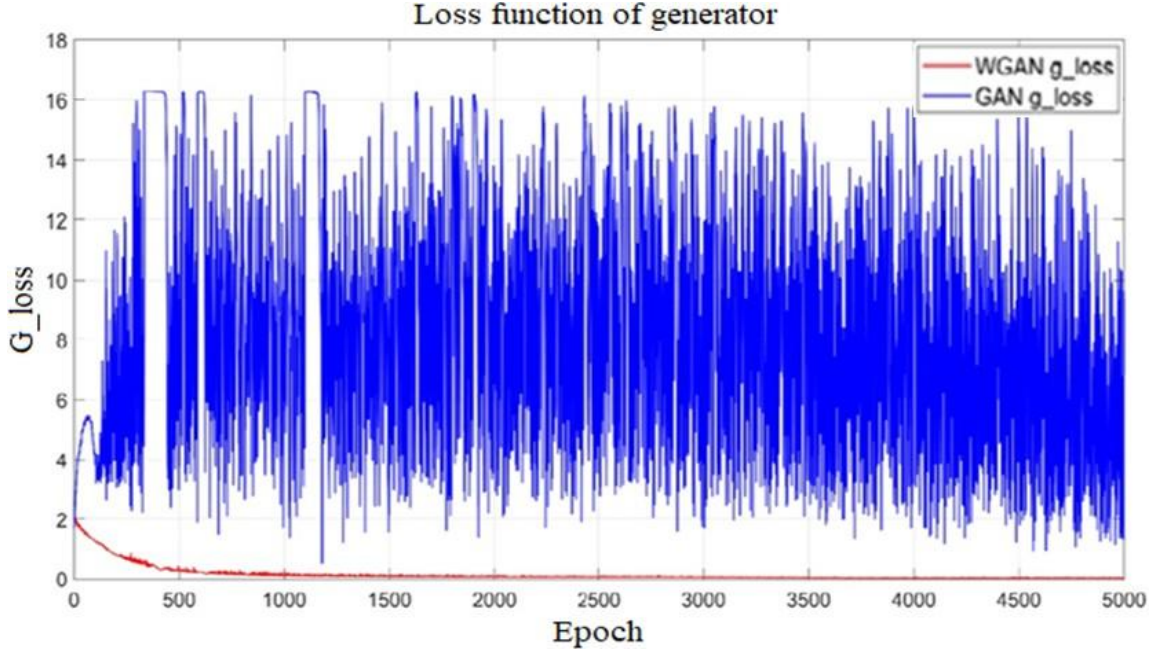

Figure 5: Generator loss functions for GAN (blue) and WGAN (red) over 5000 training epochs.

*3.3. Generation and Stability Analysis of V-O Compositions*

Our WGAN framework successfully generated a diverse set of V-O compositions commonly used in electronic devices: 184 $VO_2$ compositions, 152 $V_2O_3$ compositions, and 115 $V_2O_5$ compositions. These structures were generated in latent space and subsequently reconstructed to real space using our trained VAE, enabling comprehensive DFT analysis of their properties.

We established more rigorous stability criteria compared to previous work [28], which considered materials stable at formation energies $E_f \leq 0.5$ eV/atom. Our refined classification defines stability through two key metrics: formation energy and distance to the convex hull. We classify a structure as stable when its formation energy is negative and its distance to the convex hull is $\leq 300$ meV/atom. Meta-stable structures are defined by negative formation energy and convex hull distance $\leq 500$ meV/atom. This stricter definition ensures higher confidence in the practical viability of identified structures.

The formation energy distributions across the three compositions, shown in Fig. 6, reveal distinct stability patterns. $VO_2$ and $V_2O_3$ compositions demonstrate concentrated stability profiles, suggesting more consistent structural patterns, while $V_2O_5$ shows a broader distribution, indicating a wider exploration of metastable configurations.

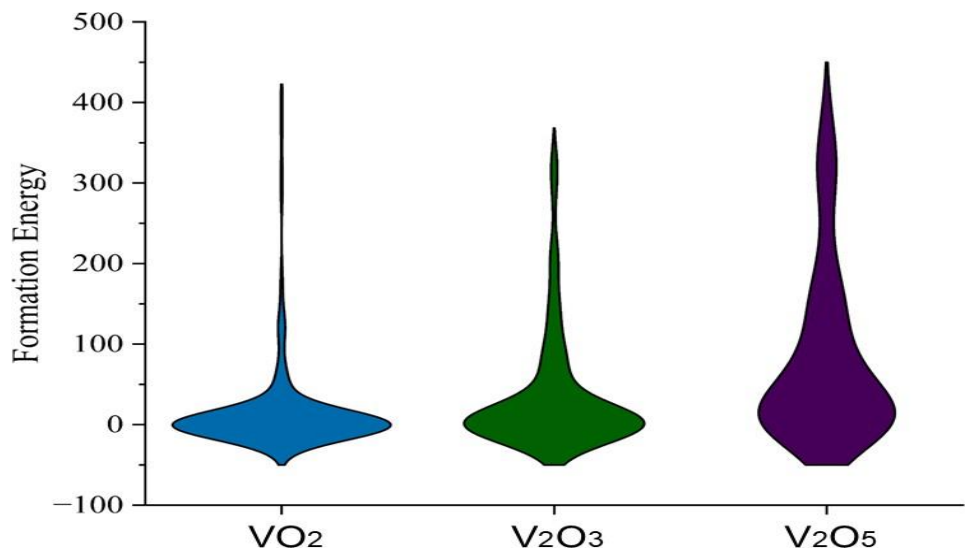

Figure 6: The distribution of formation energies for the generated $VO_2$, $V_2O_3$, and $V_2O_5$ compositions.

### 3.4. DFT Calculation and Thermodynamic Stability

For $VO_2$ compositions, our model achieved remarkable success rates: of the 184 generated structures, 124 met the stability criteria from [28], and 119 newly generated materials exhibited negative formation energies. Under our stricter criteria, 56 structures qualified as stable and 25 as meta-stable. Fig. 7 visualizes these formation energies relative to the Materials Project convex hull, demonstrating our model's ability to generate thermodynamically favorable configurations.

The $V_2O_3$ system yielded particularly significant results. From 152 generated structures,

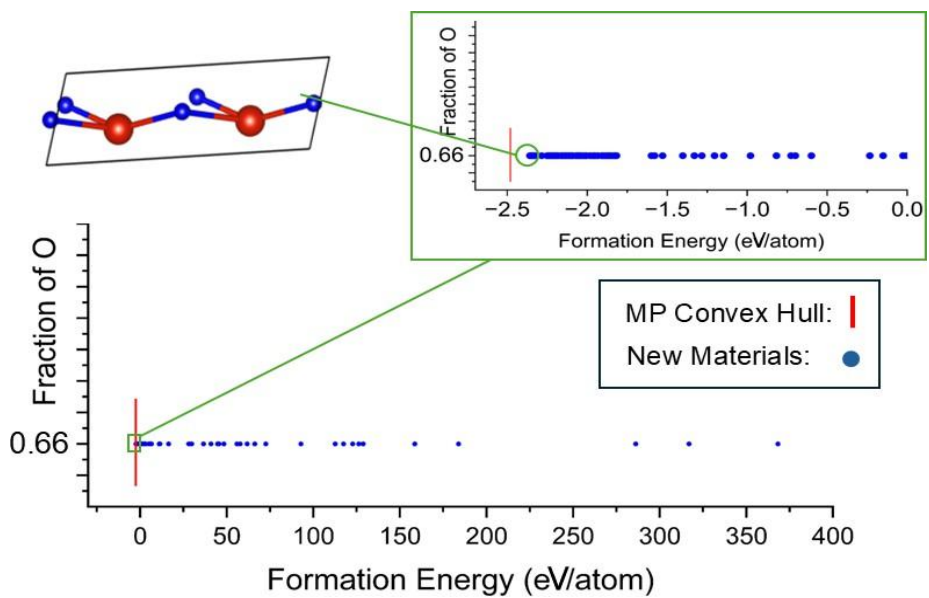

Figure 7: Formation energies of the generated $VO_2$ compositions.

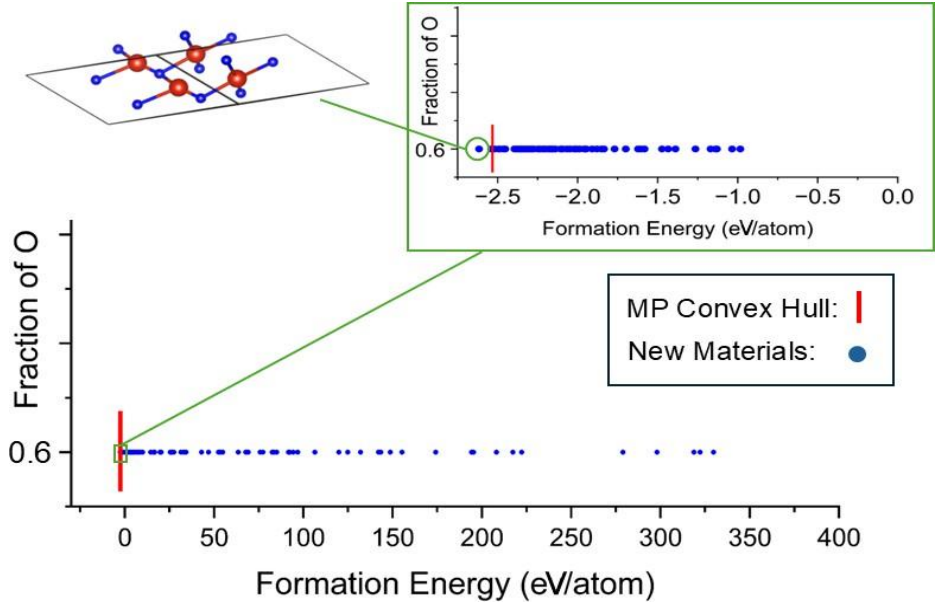

Figure 8: Formation energies of the generated $V_2O_3$ compositions.

80 satisfied the original stability criteria [28], with 77 showing negative formation energies. Using our enhanced criteria, 33 structures were classified as stable and 18 as meta-stable. Most notably, as shown in Fig. 8, we identified two $V_2O_3$ configurations with formation energies below the Materials Project convex hull, representing potentially new, thermodynamically favorable phases not previously documented.

For $V_2O_5$ compositions, we generated 115 structures, with 34 showing negative formation energies. Under our stringent criteria, two structures qualified as stable and one as meta-stable, as illustrated in Fig. 9. The complete distribution of stability metrics across all V-O compositions is summarized in Table 2.

## 3.5. Phonon Calculations and Dynamic Stability

Phonon calculations offer critical insights into a material's vibrational properties and potential structural behavior. In this work, we computed the phonon dispersion relations for the V–O compositions with the lowest formation energies to assess their dynamic stability. The resulting phonon spectra illuminate key aspects of lattice vibrations, interatomic interactions, and overall structural stability.

Table 2: Summary of the generated V-O compositions, including the total number of materials, the number of stable and meta-stable structures.

| Materials | Total | Stable in [28] | Negative | Stability | Meta-stability |
|---|---|---|---|---|---|
| $VO_2$ | 184 | 124 (67.4%) | 119 (64.7%) | 56 (30.4%) | 25 (13.5%) |
| $V_2O_3$ | 152 | 80 (52.6%) | 77 (50.6%) | 33 (21.7%) | 18 (11.8%) |
| $V_2O_5$ | 115 | 34 (29.6%) | 34 (29.6%) | 2 (1.7%) | 1 (0.87%) |
| All Materials | 451 | 238 (52.8%) | 230 (51%) | 91 (20.2%) | 44 (9.75%) |

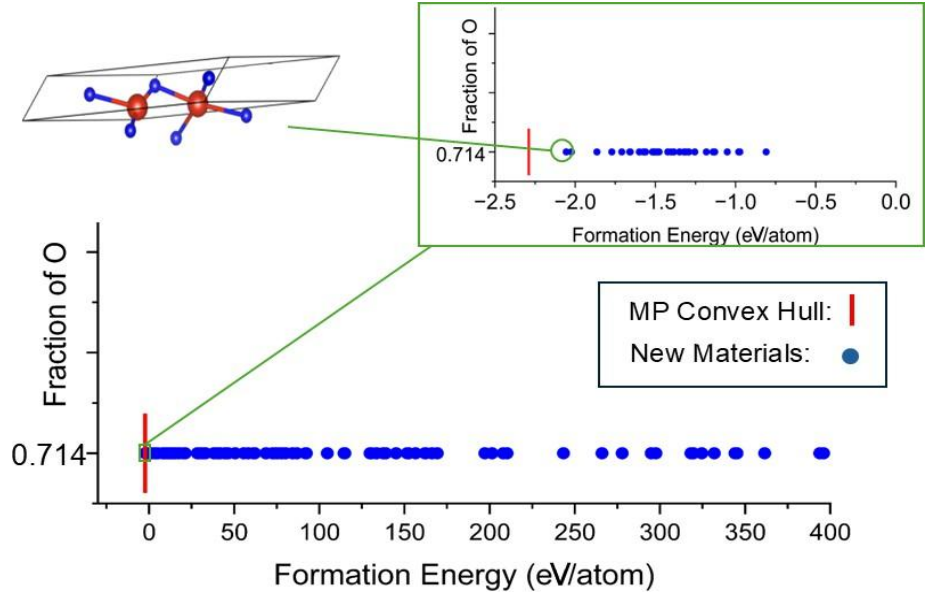

Figure 9: Formation energies of the generated $V_2O_5$ compositions.

Fig. 10 presents the phonon dispersion curves and phonon DOS for selected $VO_2$ (see Fig. 7 for its structure). A few shallow imaginary branches are evident, consistent with the known 0 K behavior of the rutile phase of $VO_2$, which is marginally unstable and naturally shifts to a monoclinic distortion at lower temperatures [44]. The slight low-frequency modes persisting at full convergence align with the well-documented phase transition in $VO_2$: the rutile phase is favored above ∼ 340 K, whereas a monoclinic form emerges below that temperature [45]. Consequently, small imaginary frequencies at 0 K do not preclude the real-world stability, as the rutile phase of $VO_2$ is often observed under ambient or elevated temperatures. Moreover, our formation energy calculations place this structure near the convex hull, reinforcing that it is thermodynamically viable and likely at least metastable—if not fully stable—under typical synthesis conditions. Additional details about how we selected supercell sizes and how these choices affect the phonon dispersion are available in the Supplementary Information (Figs. S1 and S2).

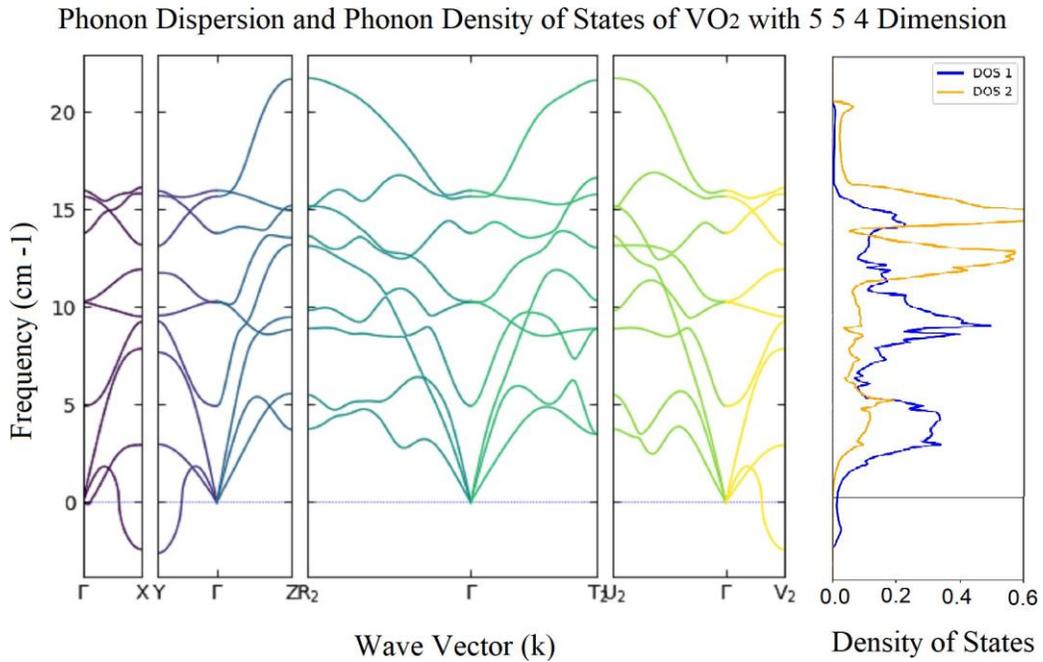

Figure 10: Phonon dispersion and phonon density of states (DOS) of $VO_2$ computed for a $5 \times 5 \times 4$ supercell.

Fig. 11 presents the phonon-dispersion curves and phonon DOS for selected $V_2O_3$, (see Fig. 8 for its geometry). Although most phonon branches remain well above zero frequency, a few minor imaginary modes are visible in certain regions of the Brillouin zone. Such soft modes do not necessarily imply a fundamental dynamic instability, especially given that larger supercells would likely reduce or eliminate them. Here, a 3×3×2 supercell was used to manage computational costs, so any remaining imaginary branches may stem from finite-size constraints, incomplete long-range force sampling, or subtle convergence parameters—rather than a genuine structural instability. Our similar analysis of $VO_2$ (Supplementary Information) supports this interpretation. The formation energy of this $V_2O_3$ phase is close to the convex hull, indicating that it is energetically viable and likely metastable. In practical syntheses, finite temperatures and pressures often stabilize phases that appear partially unstable at 0 K [46]. Consequently, while minor imaginary modes appear in the $3 \times 3 \times 2$ phonon calculation, their limited magnitude and the broader thermodynamic context suggest that the selected $V_2O_3$ phase remains a strong candidate for experimental realization.

Fig. 12 shows the phonon dispersion and phonon DOS for the selected $V_2O_5$ structure (its geometry is depicted in Fig. 9). Similar to $V_2O_3$ composition, two minor imaginary

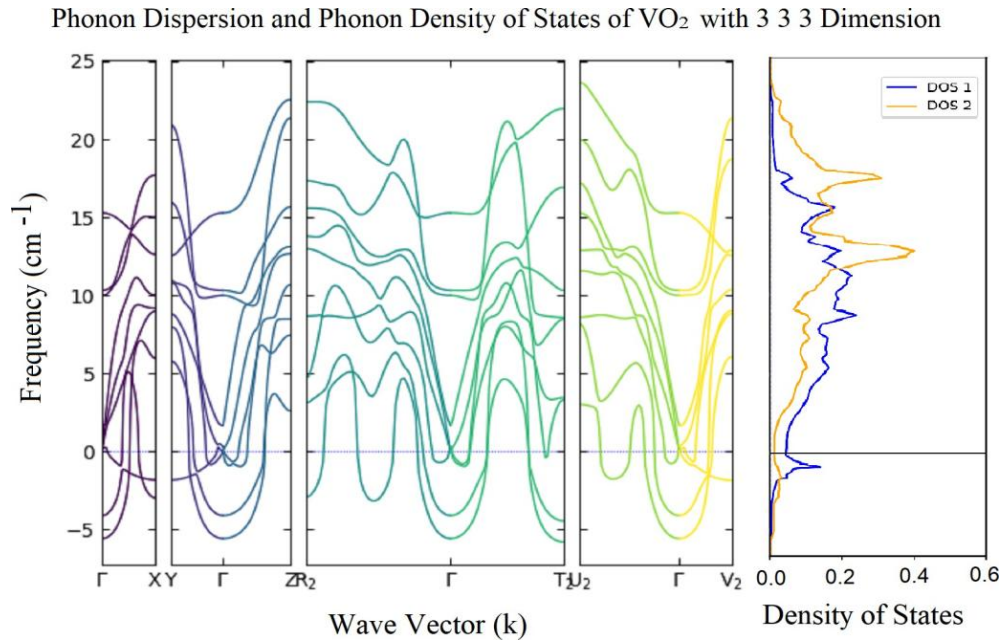

Figure 11: Phonon dispersion and phonon DOS of selected $V_2O_3$ computed for a $3 \times 3 \times 2$ supercell.

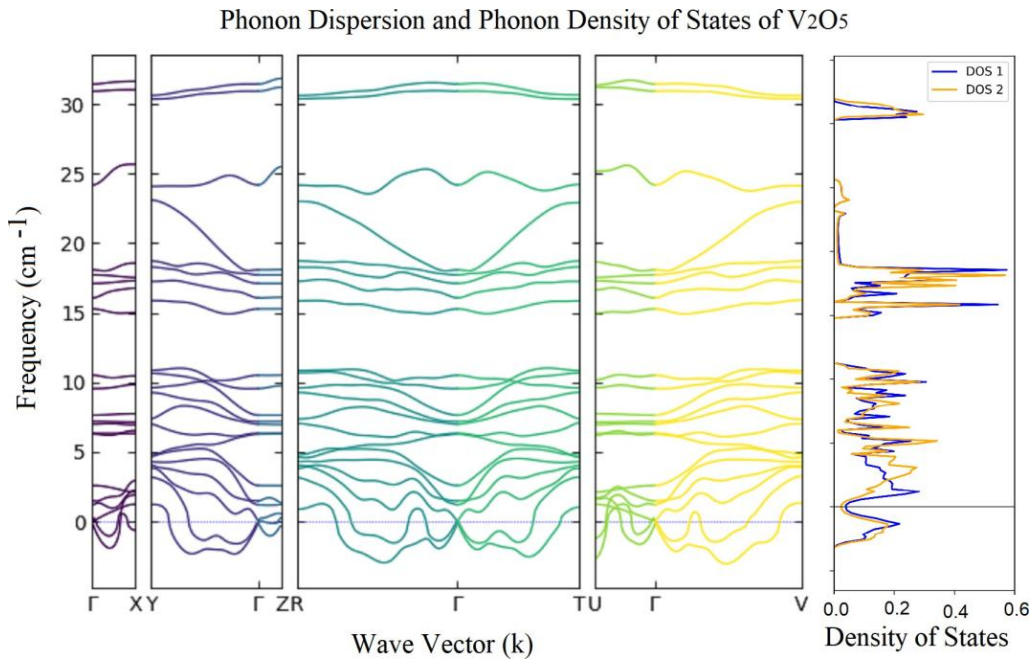

Figure 12: Phonon dispersion and phonon DOS of selected $V_2O_5$ computed for a $3 \times 2 \times 2$ supercell.

modes—likely due to the $3 \times 2 \times 2$ supercell's finite-size constraints rather than an intrinsic instability. Because their magnitudes are minimal and the formation energy lies near the known stable phases, these modes do not invalidate the phase's viability. While larger supercells could reduce these artifacts further, the associated computational cost becomes a limiting factor. Taken together, the minor imaginary frequencies, in conjunction with a near-hull formation energy, indicate that this $V_2O_5$ composition is plausibly stable or metastable and thus merits further investigation.

### 3.6. Electronic Property Analysis

To comprehensively evaluate the potential of our generated materials for electronic applications, we conducted detailed electronic structure calculations on the most thermodynamically stable configurations identified for each composition ($VO_2$, $V_2O_3$, and $V_2O_5$). Our analysis employed spin-polarized DFT+U calculations with the Hubbard U parameter set to 3.25 eV for the V 3d states [47], following established protocols for accurately describing electron correlations in vanadium oxides.

For the generated $VO_2$ structure with the lowest formation energy (see Fig. 7 for its atomic structure and energy position on the convex hull, and Fig. 10 for its phonon dispersion), our calculations revealed intriguing electronic behavior characterized by strong spin-dependent properties. Figure 13 shows the spin-polarized band structure and density of states (DOS) calculations indicating a marked asymmetry between spin channels. The spin-down channel (red lines) exhibits a substantial indirect bandgap of approximately 2.7 eV, indicating insulating behavior. In contrast, the spin-up channel (blue lines) maintains a nearly metallic character with only a minimal gap, near the Fermi level (set at 0 eV in Vaspkit). This pronounced spin dichotomy manifests clearly in the DOS, where the spin-down states show a distinct gap around the Fermi level while spin-up states maintain finite density. Such electronic structure characteristics strongly suggest half-metallic behavior, where one spin channel provides conduction pathways while the other remains insulating. The distinct spin-dependent bandgaps make this material particularly promising for spin-filtered electronics and quantum computing applications, where controlled manipulation of electron spin states is crucial.

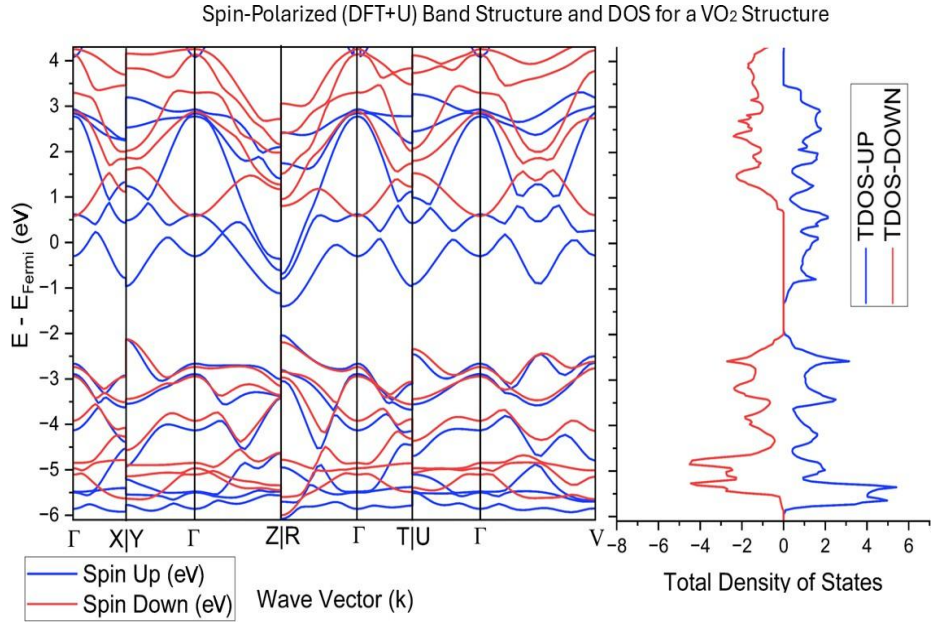

Figure 13: Spin-polarized DFT+U band structure (left) and spin-resolved DOS (right) for a triclinic $VO_2$ phase.

The generated $V_2O_3$ composition (see Fig. 8 for its atomic structure and energy position on the convex hull, and Fig. 11 for its phonon dispersion) displays intriguing electronic properties as shown in Fig. 14. This $V_2O_3$ structure has formation energy below the Materials Project convex hull, and exhibits electronic characteristics that parallel those of $VO_2$ but with distinct quantitative differences. The spin-resolved calculations again reveal asymmetric behavior between spin channels, indicative of potential half-metallic properties. The spin-down channel has a band gap of 3.73 eV wider than the $VO_2$ structure. The band structure shows complex hybridization between V 3d and O 2p states across a broad energy range that can be important in determining the material's electronic transport properties and magnetic behavior. These electronic structure calculations provide valuable insights into the potential applications of our generated materials. The half-metallic behavior observed in $VO_2$ and $V_2O_3$ structures suggests promising applications in spintronics and magnetic devices, where spin-polarized current is desirable.

For generated $V_2O_5$ (see Fig. 9 for its atomic structure and energy position on the convex hull, and Fig. 12 for its phonon dispersion), the electronic structure analysis, in Fig. 15, reveals that the spin-polarized calculations are densely packed, highly entangled electronic states near the Fermi level in both spin channels. Unlike $VO_2$ and $V_2O_3$, we observe no clear spin-selective gap formation, even with the inclusion of the Hubbard U term. The DOS calculations indicate the presence of accessible states at and around the Fermi level for both spin orientations, suggesting that electrons in both spin channels can participate in conduction processes. This electronic structure does not support half-metallic behavior but instead points to a more conventional metallic character with possible weak magnetic ordering. These characteristics make this triclinic $V_2O_5$ composition particularly promising for transparent conducting oxides, high-performance electrochromic devices, and energy storage systems where rapid electron transport in both spin channels is advantageous.

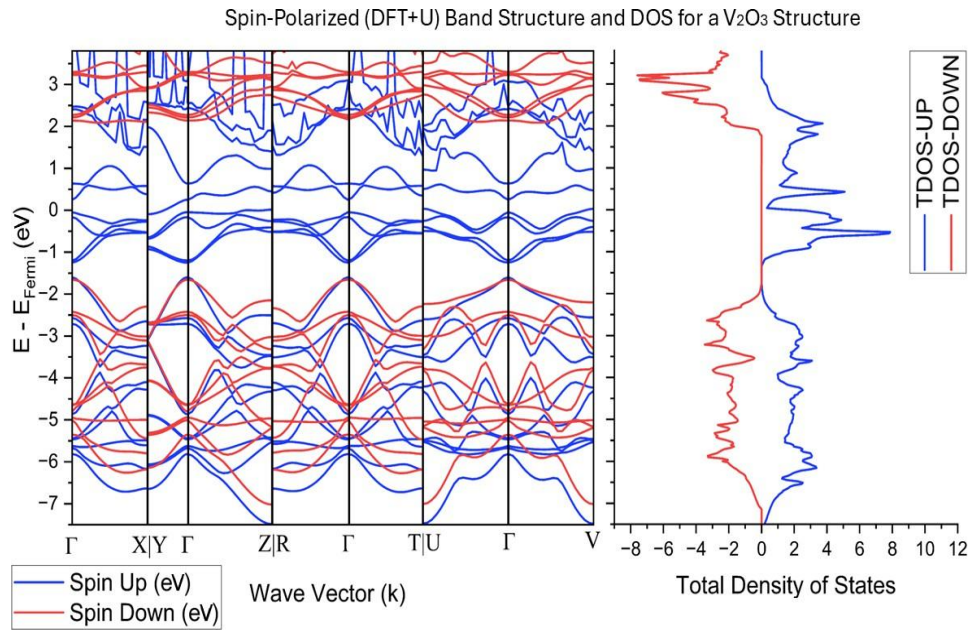

Figure 14: The triclinic $V_2O_3$ phase is represented with the spin-polarized DFT+U band structure and the spin-resolved DOS.

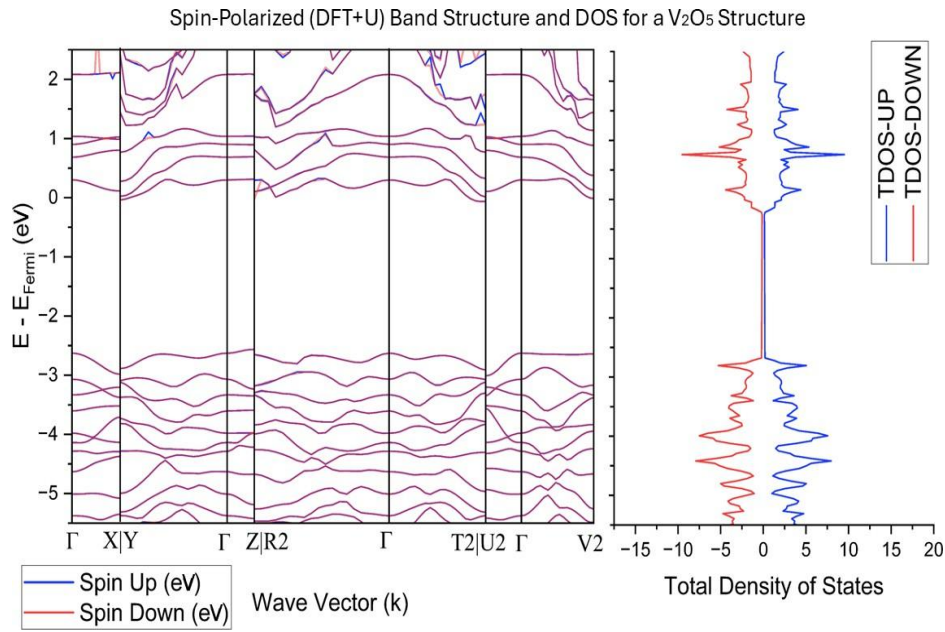

Figure 15: Spin-polarized DFT+U band structure and spin-resolved DOS for the triclinic $V_2O_5$ phase.

The low symmetry of these triclinic structures introduces additional complexity in the band structures, manifesting as non-smooth features and intricate band crossings. This complexity arises from the reduced symmetry of the Brillouin zone and the resulting increase in allowed band mixing. The structural distortions in these materials appear to play a significant role in determining their electronic properties, particularly affecting the bandwidth and band hybridization characteristics. The presence of strong electron correlations, evidenced by the significant impact of the Hubbard U term on the electronic structure, indicates that these materials may exhibit interesting phenomena under external stimuli such as temperature, pressure, or electric fields. This susceptibility to external perturbation could be particularly valuable for sensing applications or switchable electronic devices.

## 4. Conclusion

In this study, we present an innovative inverse design framework that uses enhanced generative AI architectures to produce stable vanadium oxide compositions, showing that properly constrained machine learning models can navigate complex chemical spaces and accelerate the discovery of novel functional materials. We develop a WGAN with stability constraints for smoother training and improved generation quality, coupled with a voxel-based VAE to capture atomic positions and lattice parameters while maintaining chemical validity. The framework's effectiveness is demonstrated by generating and characterizing 451 unique V–O compositions, of which 91 are stable under strict criteria and 44 are metastable, surpassing previous benchmarks. Spin-polarized DFT+U calculations confirm that the generated materials exhibit diverse and technologically relevant properties. Notably, several $V_2O_3$ configurations have formation energies below the Materials Project convex hull, illustrating the framework's ability to discover materials beyond known phase spaces. The new $VO_2$ structures show half-metallic behavior with spin-dependent bandgaps, while $V_2O_5$ structures exhibit metallic character with unique electronic transport properties. Furthermore, phonon-dispersion calculations for selected $VO_2$, $V_2O_3$, and $V_2O_5$ phases support their dynamic stability and meta-stability. Although minor imaginary modes are observed in some cases at 0 K, these typically stem from finite-size limitations or known phase transitions (e.g., rutile $VO_2$), thereby likely indicating the viability of these materials under practical synthesis and operating conditions.

Beyond vanadium oxides, this framework can be adapted to other materials by adjusting chemical constraints and stability criteria, especially for transition metal oxides. Future advances could integrate additional property predictions into the generative process to target materials with specific electronic, optical, or magnetic features. The newly identified stable structures, particularly the novel $V_2O_3$ phases, merit experimental synthesis and characterization. Further optimization of the WGAN architecture could improve generation speed and scalability to larger chemical spaces. By combining artificial intelligence with materials science, this approach establishes a robust methodology for accelerating the identification and development of next-generation electronic materials and quantum technologies.

## Data Availability

All data generated in this study, including the materials in VASP format and calculation guidance, are available in the GitHub repository: https://github.com/INQUIRELAB/WGAN-VAE-V-O-Stable-Compositions. The repository contains the complete set of crystal structures, computational parameters, and analysis scripts used in this work.